# Investigation of Vortex Structures in Gas-Discharge Nonneutral Electron Plasma: I. Experimental Technique

N. A. Kervalishvili

Iv. Javakhishvili Tbilisi State University, E. Andronikashvili Institute of Physics
Tbilisi 0177, Georgia.  <n_kerv@yahoo.com>

**Abstract.** The nonperturbing experimental methods have been described, by means of which the solitary vortex structures in gas-discharge nonneutral electron plasma were detected and investigated. The comparison with the experimental methods used in devices with pure electron plasma was made. The problems of shielding the electrostatic perturbations in nonneutral plasmas were considered.

## I. Introduction

Nonneutral plasma is a peculiar kind of medium differing from all well-known states of matter: plasma, gas, liquid and solid. Nonneutral plasmas consist of charged particles only (or predominantly) of one sign due to which they are characterized by large electric fields. In nonneutral plasmas the predominant type of interaction of charged particles is electrostatic repulsive forces. Therefore, the laboratory nonneutral plasmas can be confined only by a strong magnetic field. For this purpose, the devices with crossed electric and magnetic fields are used. Nonneutral electron, ion, and positron plasmas are created by filling such devices with charged particles. But, one of the most simple and efficient ways of obtaining and studying the nonneutral electron plasma is the use of discharge in crossed electric and magnetic fields. In the simplest case, the discharge device is two coaxial cylindrical electrodes located in longitudinal magnetic field. One of electrodes serves as an anode (external in case of magnetron or internal in case of inverted magnetron) and the second one – as a cathode. Along the magnetic field, the discharge space between the cylinders is limited by end electrodes, which are under the cathode potential. The source of electrons and ions in the discharge is the ionization of neutral gas atoms by electrons. The parameters of the discharge are such that the ions are not magnetized and leave the discharge gap without collisions. At the same time, the electrons are strongly magnetized and are trapped by magnetic field. Under such conditions the nonneutral electron plasma in the discharge is formed. The sheath of nonneutral electron plasma is located near the anode surface and the whole discharge voltage falls on it [1-8].

The large electric fields and the absence of charges of opposite sign are one of the main reasons of the different approach to the problems of shielding the electrostatic perturbations and to the use of Langmuir probes for local diagnostics in nonneutral plasmas. These problems are dealt with in Sec. II of the given paper. The following sections (III-VI) are devoted to the description of nonperturbing and contactless experimental methods that allow to detect in discharge electron sheath the solitary vortex structures with high electron density, to study their parameters and the processes of formation, evolution and dynamics [9-12]. In Sec. VII the comparison is given with the other experimental methods used for studying the vortex structures in pure electron plasma confined in Penning-Malmberg traps and in toroidal systems.

## II. The problem of shielding in nonneutral plasmas

The specific peculiarity of conducting medium such as quasi-neutral plasma and electrolytes is the capability to shield the electrostatic perturbations caused by external electric fields (test charge, probe, walls). The process of shielding consists in accumulation of the charges of opposite sign around the test charge, and their fields compensate the field of the test charge at the characteristic distance, the so-called Debye shielding length. Though the Debye shielding has been known for a sufficiently long time, the interest to this problem is still actual because of the peculiarities of shielding in the presence of collisions, of magnetic field, of secondary emission from the wall or from the probe, and of the dynamics of applied perturbing potential [13-16]. It should be noted that in all these cases we have two signs of charges in neutral conducting medium. However, a peculiar interest for the problem of shielding is the conducting medium consisting of charged particles of only one sign. Because of the large repulsive forces such medium cannot be in liquid or solid state at any temperature. This medium is called a nonneutral plasma. A main problem of shielding in this plasma is the absence of charges of opposite sign and the presence of large internal electric fields. Besides, the electric field is not constant, it increases continuously to the direction perpendicular to the magnetic field. Therefore, such plasma cannot be considered to be the unbounded medium under any conditions.

As for the shielding in nonneutral plasmas there is not a well-established picture. It is generally considered that the exponential Debye shielding takes place as well in nonneutral plasmas. But there are papers [17], in which it is shown that in pure electron plasma the negative potentials are not shielded as efficiently as the positive potentials. Besides, the negative potentials can penetrate deeply into plasma and exert a strong influence on the motion of electrons. It is not our purpose to review the papers on the problems of shielding in nonneutral plasmas. However, we are going to consider some problems from purely physical point of view:

1. The difference between the positive and negative test charges in nonneutral electron plasma is that the excess of electrons are accumulated around the positive charge and we have two types of charges as in the case of quasi-neutral plasma. Around the negative charge the electron deficiency is formed and we can speak about the equalizing (smoothing) of electric fields at large distances rather than about the shielding (compensation).

2. Let us describe the Debye sphere around the test charge. In quasi-neutral plasma this sphere is equipotential and the electric field on it is equal to zero. In nonneutral electron plasma such sphere cannot be equipotential and the electric field on its surface is different at different points.

3. The sheath of nonneutral electron plasma can be considered as a shielding sheath near the anode. On the other hand, the positive charge on the anode surface can be considered as a charge induced by the electron sheath. In this case, the perturbation inside the sheath caused by the excess of electrons leads to the appearance of the additional induced charge on the anode surface. Hence, the perturbation inside the electron sheath caused by the excess of electrons can be stretched to the sheath boundaries and further up to the anode surface and maybe to the cathode surface depending on the value of perturbation.

4. Generally speaking, the interaction of particles in the nonneutral plasma differs so strongly from the interaction of particles in the quasi-neutral plasma that it can be rather compared to the gravitational interaction. At the gravitational interaction all the particles are attracted to each other. At the electrostatic interaction of the charged particles of one sign all the particles are repulsed from each other. In both cases, the long-range action takes place, and the force of interaction depends on the distance and the number of particles. Probably, there is a certain analogue as well in the problems of shielding between the gravitational field and the electric field of the charges of one sign.

The problem of shielding is tightly connected with one of the most commonly used efficient methods of local diagnostics of low-density plasma – by electric probes. The problem of shielding in nonneutral electron plasma is reduced to the fact that Langmuir probes will introduce the strong perturbations into plasma sheath. This fact is mentioned by many authors [17-19]. However, in the cases, when the other diagnostics cannot be applied, the emissive Langmuir probes are used but with some restrictions [20-23].

We refused to use the Langmuir probes at the early stage of investigations of the discharge in crossed electric and magnetic fields [7,8], when the experiment showed that a small Langmuir probe has a strong influence on the whole plasma sheath. Figure 1 shows the photo of electron sheath in inverted magnetron and the dependences of discharge current and of floating potential of Langmuir probe on the distance between the probe and the anode surface.

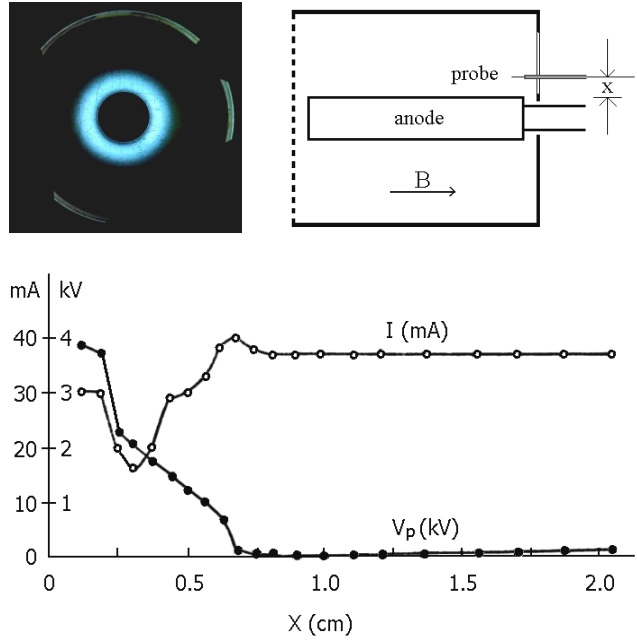

Fig.1. Nonneutral electron plasma sheath in the geometry of inverted magnetron.
$r_a = 0.9 cm$; $r_c = 3.7 cm$; $L = 7 cm$; $B = 1 kG$; $V = 4 kV$; $p = 3\times 10^{-4} Torr$.

Here, $r_a$ is the anode radius, $r_c$ is the cathode radius, $L$ is the anode length, $B$ is the magnetic field, $V$ is the discharge voltage, $p$ is the neutral gas pressure, $I$ is the discharge current and $V_p$ is the floating potential of probe. The probe was located along the magnetic field and was connected to the electrostatic voltmeter in order to be at the floating potential. The probe shaft was covered with ceramic tube.

From the picture it is seen that despite the fact that the current on the probe equals zero, the discharge current changes more than twice at the displacement of probe across the electron sheath. It turned out that the introduction of any outside body, be it a conductor or a dielectric, into the electron sheath has a strong perturbing action. Therefore, for studying the inhomogeneities in gas-discharge nonneutral electron plasma the preference has been given to wall probes.

**III. The method of two wall probes**

The inhomogeneity in electron sheath means the local increase or decrease of electron density inside the electron sheath and hence, the local increase or decrease of electric field in the sheath. This process is accompanied with the local increase or decrease of the induced charge on the anode surface. The orbital motion of inhomoheneity is accompanied with the motion of the induced charge on the anode surface. The wall probe is a small part of electrode connected with that electrode by the resistance. The value of resistance has a significant importance depending on the function of probe. If the wall probe is used for excitation of oscillations in plasma, the resistance should be relatively large in order the probe field to have action on plasma. If the probe is used for diagnosis of inhomogeneities, the resistance should be sufficiently small in order to neglect the influence of the probe on plasma. On the right side of Fig. 2 the circuit of wall probe is given.

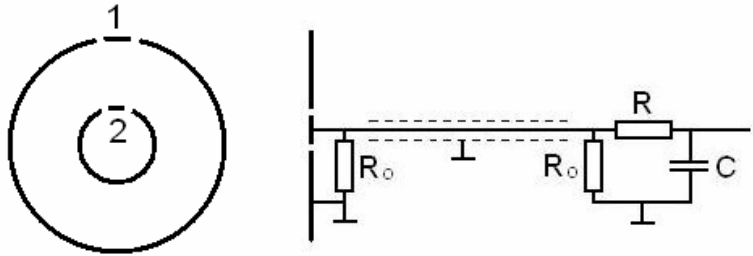

Fig. 2 The circuit of wall probe

The signal from the wall probe allows to measure directly the amplitude of oscillations of the electric field on the anode surface ($\Delta E$) during the passing of ingomogeneity by the probe. As the wall probe registers simultaneously the electric field of the induced charge and the current from the plasma, it is necessary to determine the contribution of each of them to the total signal from the probe. The current on the probe can be measured independently. For this purpose, the probe (collector) should be shielded putting it behind the slit or the holes in the electrode.

Generally speaking, the wall probe is ideally suitable just for the nonneutral plasma, as it registers the moving charged inhomogeneity. However, it is difficult to obtain a detail information about this inhomogeneity, since the amplitude of signal on the wall probe depends on the inhomogeneity charge, on the distance between the inhomogeneity and the probe and on the shape of inhomogeneity itself.

To determine the charge and the trajectory of inhomogeneity, the method of two wall probes was developed [9]. In this method, the signals from the wall probe of cylindrical anode and from the wall probe of cylindrical cathode are measured simultaneously. The probes are located on the same azimuth (left side of Fig. 2). If the transverse dimension of inhomogeneity is small as compared to the distance between the inhomogeneity and each of the probes, and besides, the inhomogeneity is stretched along the magnetic field, the value of signal from the moving inhomogeneity on each of the probes will depend only on linear charge density of inhomogeneity and on the distance between the inhomogeneity and the corresponding probe. Therefore, measuring simultaneously the amplitudes of the signals from two probes, one can determine the linear charge density and the position of inhomogeneity.

Let us consider two coaxial cylindrical electrodes with a thin charged filament located between the cylinders parallel to the axis of cylinders. The filament rotates around the axis of cylinders with the angular velocity $\omega$. The coordinates of filament are: $r = r_0$ and $\theta = \omega t$, where $r$ is the radius, and $t$ is the time. The electric fields at the fixed points of cylindrical surfaces (in the places of the location of probes), caused by the rotating charged filament have the following form:

$$E_1 = -\frac{4q}{r_1}\left\{\sum_{m=1}^{\infty}\frac{r_0^{2m}-r_2^{2m}}{r_1^{2m}-r_2^{2m}}\left(\frac{r_1}{r_0}\right)^m \cos m\omega t + \frac{1}{2}\frac{\ln(r_2/r_0)}{\ln(r_2/r_1)}\right\}$$

$$E_2 = \frac{4q}{r_2}\left\{\sum_{m=1}^{\infty}\frac{r_1^{2m}-r_0^{2m}}{r_1^{2m}-r_2^{2m}}\left(\frac{r_2}{r_0}\right)^m \cos m\omega t + \frac{1}{2}\frac{\ln(r_0/r_1)}{\ln(r_2/r_1)}\right\}$$

Here, $E_1$ and $E_2$ are the electric fields on the surfaces of external and internal cylinders, $r_1$ and $r_2$ are the radii of external and internal cylinders, respectively, $q$ is the charge of the unity of filament length and $r_0$ is the radius of drift orbit of filament. The amplitudes of electric field oscillations on the probes at the rotation of filament about the axis of discharge device $\Delta E = E(\omega t = 0) - E(\omega t = \pi)$ will be equal respectively to:

$$\Delta E_1 = -\frac{4q}{r_1}\sum_{m=1}^{\infty}\frac{r_0^{2m}-r_2^{2m}}{r_1^{2m}-r_2^{2m}}\left(\frac{r_1}{r_0}\right)^m (1-\cos m\pi)$$

$$\Delta E_2 = \frac{4q}{r_2}\sum_{m=1}^{\infty}\frac{r_1^{2m}-r_0^{2m}}{r_1^{2m}-r_2^{2m}}\left(\frac{r_2}{r_0}\right)^m (1-\cos m\pi)$$

By measuring simultaneously $\Delta E_1$ and $\Delta E_2$ we can determine $r_0$ and $q$. The output signals from wall probes are extracted by the matched rf cables, are integrated, and are registered by single-scan oscillographs. As the inhomogeneity drifts about the axis of the discharge device, the signals are periodically repeated and we can follow, within the scanning time, how the charge and the trajectory of inhomogeneity are changed in time. Besides, the type (clump or hole) and the shape of the inhomogeneity can be estimated from the polarity and the shape of the signals, respectively.

This method was used as well for determination of charges and trajectories in the case of two inhomogeneities moving with different angular velocities [9]. For this purpose, it is sufficient to measure the total amplitudes of electric fields on the probes at the maximum approach of inhomogeneities and the difference in amplitudes at their maximum removal from each other.

Figure 3 shows the typical oscillograms of signals (oscillations of electric fields) from the probes located on internal and external electrodes at moving one (A) and two (B) inhomogeneities about the axis of discharge device.

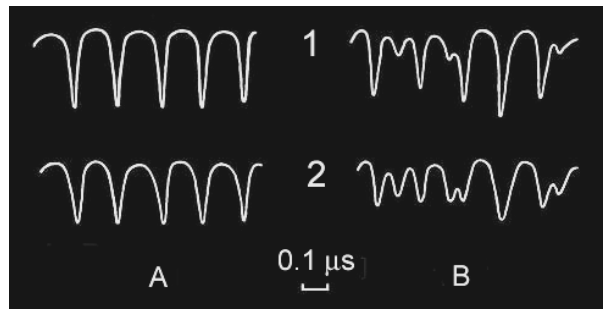

Fig.3. Signals from wall probes in magnetron geometry.
$r_a = 3.2 cm$; $r_c = 1.0 cm$; $L = 7 cm$; $B = 1.9 kG$; $V = 1.5 kV$; A – $p = 1\times10^{-5}$; B – $1\times10^{-4} Torr$

The method of two wall probes was developed for the geometries of magnetron and inverted magnetron having the internal and external cylindrical electrodes. In the case of Penning cell, in which the internal cylindrical cathode is absent, the use of only one anode wall probe will be enough if we use the other methods for determination of $r_0$ [10].

**IV. The method of electron ejection analysis**

In the process of investigation of inhomogeneities, it appeared that they are the compact vortex structures with the high electron density stretched along the magnetic field [9]. Besides, it turned out that the presence of vortex structures in the discharge electron sheath is accompanied with the ejection of electrons along the magnetic field to the end cathodes, both from the vortex structures and from the regions of electron sheath adjacent to them [9]. This current is the periodically repeated pulses appearing at the formation of vortex structures, at their approach and at the radial displacement of the vortex structure towards the cylindrical cathode [9-11]. Besides, the continuous electron current goes to the end cathodes from the vortex structures and the regions of electron sheath adjacent to them [24]. This phenomenon was used for determination of the cross-section of inhomogeneity. As the electron density inside the vortex structure is much higher than the density of sheath electrons and, correspondingly, the current of electrons from the vortex structure is much intense than the current of electrons form the adjacent region of electron sheath, the size of region of the intense ejection is identified with the size of inhomogeneity.

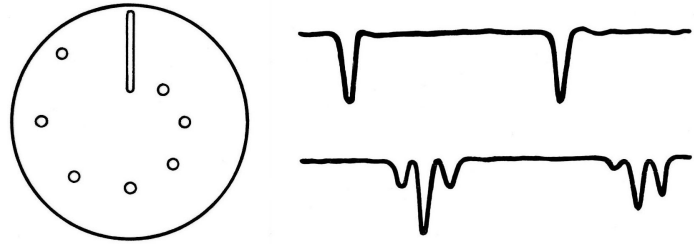

Fig.4. Endplate cathode with probes: upper oscillogram is the electron current through the slit and lower oscillogram is the electron current through the holes.

For carrying out the measurements the endplate cathode of discharge device was used having the form of a disc with a narrow radial slit and holes located along the flat spiral that is shown in Fig. 4 [12]. The observation of electron current through the slit (upper oscillogram) and of the electron current through the holes (lower oscillogram) made at one and the same time allowed to determine the azimuthal and radial dimensions of drift inhomogeneity.

The azimuthal dimension of inhomogeneity is determined by the dependence:

$$\Delta S = 2\pi r_0 \Delta t_S / T$$

where $\Delta t_S$ is the half-width of the pulse of electron current through the slit, $T$ is the period of rotation of inhomogeneity about the axis of discharge device, $r_0$ is the radius of drift orbit of inhomogeneity, which was determined either by the method of two probes, or by the delay of pulse through the holes relative to the pulse through the slit [12].

The radial dimension of inhomogeneity is determined by the expression:

$$\Delta R = \delta \Delta t_R / T$$

Here $\delta$ is the pitch of spiral, $\Delta t_R$ is the half-width of electron current through the holes. The pulse through the holes is observed in the form of a pulse group and, therefore, one should use their envelope.

The simultaneous measurement of the transverse dimensions and of the charge of inhomogeneity allowed to determine its average electron density. For synchronous observations, the radial slit was located on the same azimuth as the wall probes. Figure 5 shows the oscillograms of the oscillations of the electric field on the anode wall probe (upper oscillogram) and of the electron current through the slit in the end cathode (lower oscilogram) for one stable vortex structure (left) and for two approaching vortex structures (right). From the figure it is seen that the region of electron ejection rotates together with the vortex structure about the axis of discharge device. At approaching the vortex structures, the pulse of electron current produced during their approach is imposed on continuous electron current from the vortex structures.

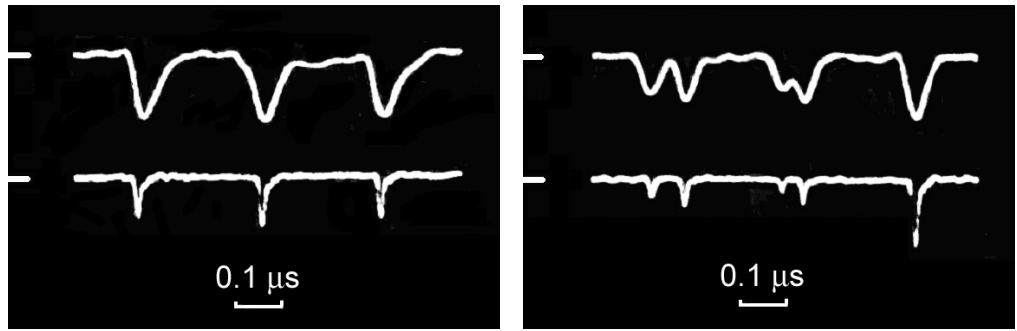

Fig.5. Continuous electron ejection from vortex structures in Penning cell
$r_a = 3.2 cm$; $L = 7 cm$; $B = 1.9 kG$; $V = 1.0 kV$; $p = 1 \times 10^{-5} Torr$

To register the pulses of electron current produced at the formation of vortex structures at their approach and at the displacement of vortex structure towards the cylindrical cathode, the second endplate cathode having the form of shielded disc was used. Figure 6 chows the schemes of probes for measuring the pulses of electron current and the oscillations of ion current in the inverted magnetron. In the same figure the typical oscillograms of signal from these probes during the periodical appearance of diocotron instability are presented. The both probes are shielded to avoid the distortion of current signal by stronger electrostatic signals from the moving inhomogeneities and oscillations of electron sheath density.

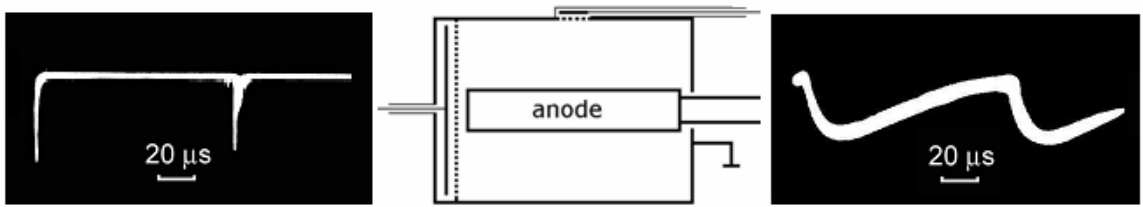

Fig.6. Probes for measuring the pulses of electron and ion currents in inverted magnetron
$r_a = 0.9 cm$; $r_c = 4 cm$; $L = 5 cm$; $V = 5 kV$; $B = 1 kG$; $p = 8 \times 10^{-5} Torr$

Figure 7 shows the oscillograms of oscillations of the electric field on the anode wall probe (upper oscillogram) and of the electron current on the disc (lower oscilogram) at formation of vortex structure in inverted magnetron (left) and at radial oscillations of vortex structure in magnetron geometry (right).
Thus, the signal from the disc gives a temporal pattern of pulse ejection of electrons along the magnetic field, and the signal from the slit determines the azimuthal dimension of the region of electron ejection. It should be noted that the electron currents on the both end cathodes are synchronous and practically identical.

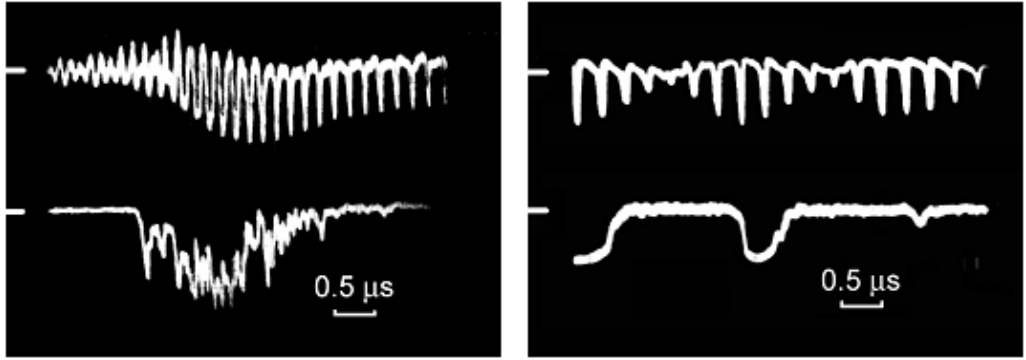

Fig.7. Formation and radial oscillations of vortex structure
Left: $r_a = 2.0 cm$; $r_c = 3.2 cm$; $L = 7 cm$; $B = 1.5 kG$; $V = 1.0 kV$; $p = 1 \times 10^{-5} Torr$.
Right: $r_a = 3.2 cm$; $r_c = 1.0 cm$; $L = 7 cm$; $B = 1.2 kG$; $V = 1.5 kV$; $p = 6 \times 10^{-6} Torr$.

**V. Contactless analogues of wall probes**

In the method of two wall probes, one of the probes is always under the high potential (several $kV$) that causes some difficulties for carrying out the experiment. To avoid this difficulty, the internal cylinder was grounded and for the external cylinder the induction method of measuring was used [25]. The essence of this method is that instead of wall probe, a longitudinal cut (a narrow slit) is made in the external cylinder along the whole length of the cylinder, and the isolated diamagnetic probe (rf cable with several turns) was wrapped around the cylinder. At the motion of inhomogeneity, the image charge induced by it overcomes the slit of the cylinder surface at the expense of the current flowing backward round the circle of the cylinder. This current flows for the whole period during which the inhomogeneity passes the slit and creates the pulse of magnetic field registered by diamagnetic probe. For comparison, in Fig.8 the signals from wall (1) and diamagnetic (2) probes are shown for the same cylindrical electrode (the cathode of inverted magnetron). The diamagnetic probe was calibrated according to the wall probe. It should be noted that the practical use of induction method is connected with definite problems. First, it is necessary to avoid the existence of any short-circuited turns in discharge device: the elements of the construction of discharge device should have cuts in corresponding places, and the cylindrical vacuum chamber should be made of high-resistive material (stainless steel or titanium). Second, the number of turns of diamagnetic probe should be chosen such that, its inductive resistance (for studied range of angular frequencies of rotation of inhomogeneities around axis of discharge device) to be much lower or much higher than the loading resistance of a probe. In the last case, it is not necessary to have an integrator and the sensitivity of the method increases.

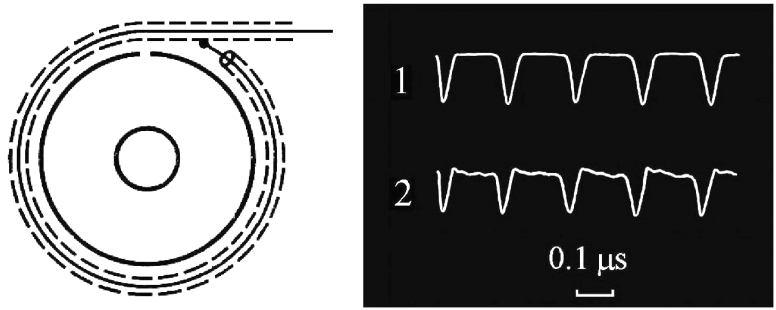

Fig.8. Signals from wall (1) and diamagnetic (2) probes
$r_a = 2.0 cm$; $r_c = 3.2 cm$; $L = 7 cm$; $B = 1.5 kG$; $V = 1.0 kV$; $p = 2 \times 10^{-5} Torr$.

The induction method has its own advantages. This is the absence of electric contact with the electrodes of discharge device; the equivalence to the very narrow wall probe the width of which is equal to the slit width; the absence of reaction at variation of the electric field of electrode and, finally, the possibility to register not only the moving inhomogeneity but also the jump of magnetic field connected e.g. with a rapid change of electron sheath density. Such measurements were made in [26] at the registration of a collapse of electron sheath in the Penning cell in the transition region of neutral gas pressures, when the ion density in the discharge electron sheath becomes comparable to the electron density. Figure 9 presents the oscillogram of the signal from the diamagnetic probe in this region of neutral gas pressures. The large positive peaks are connected with an abrupt increase of magnetic field at the collapse of electron sheath, and each peak is preceded by the signals from the inhomogeneities rotating about the axis of discharge device.

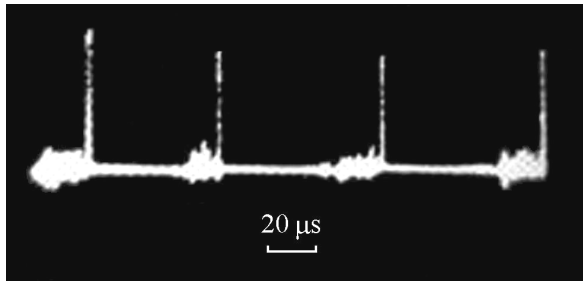

Fig.9. Signal from diamagnetic probe at the collapse of electron sheath in Penning cell.
$r_a = 3.2 cm$; $L = 7 cm$; $B = 1 kG$; $V = 1.2 kV$; $p = 1.5 \times 10^{-4} Torr$.

Let us present one more contactless method of measuring the signals from the moving inhomogeneities. This is the slit method [27] that allows to measure simultaneously the oscillations of electric fields at unlimited number of points on all electrodes of discharge device. The method consists in that the longitudinal narrow slits are made along the cylindrical electrodes but not along their whole length. The image charge, induced by inhomogeneity on the electrode surface, is moved synchronously with the motion of inhomogeneity and overcomes the slit (a cut) of the electrode surface at the expense of current flowing round the end of the slit, where a tiny isolated coil is located, registering the magnetic field of this current. Figure 10 shows the scheme of location of the wall probe and of the slit with the coil for one and the same electrode – the anode of magnetron. The same figure shows the signals from wall (upper) and slit probes. The number of coil turns (40) was chosen in order the induction resistance of the coil to be by an order of magnitude higher than the loading resistance of the probe (50 ohm).

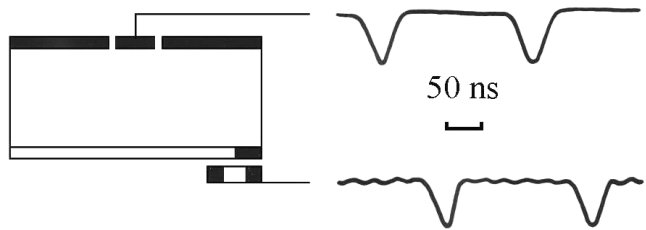

Fig.10. Anode of magnetron. Signals from wall (upper) and slit probes.
$r_a = 3.2 cm$; $r_c = 0.9 cm$; $L = 7 cm$; $B = 1.5 kG$; $V = 2.0 kV$; $p = 1.5 \times 10^{-5} Torr$.

## VI. Method of discharge interruption

Till now the main attention was paid to the nonperturbing diagnostics and to the impermissibility of the interruption of natural process of formation, evolution and interaction of vortex structures. However, interesting are not only the parameters of vortex structures but also the parameters of electron sheath itself: its dimensions, electron density, electric field on the anode, etc. Since the wall probes enable to measure the variations of electric fields and do not react to the stationary electric field, the method of a short-time interruption of discharge by applying the anode potential to one of the end cathodes was used [28]. This was done by using of a controllable miniature discharger directly in the vacuum chamber. At the same time, the discharge voltage applied between the cylindrical anode and the cylindrical cathode was maintained. The jump of electric field caused by the removal of electrons from the discharge gap was registered by the wall probe.

Figure 11 shows the oscillograms of the signals from the anode wall probe for three cases of discharge interruptions in the magnetron: 1 when there is not an inhomogeneity, 2 with a small inhomogeneity, and 3 with a great inhomogeneity (a large charge of inhomogeneity is meant). For visualization, the oscillograms in the figure are flip vertical.

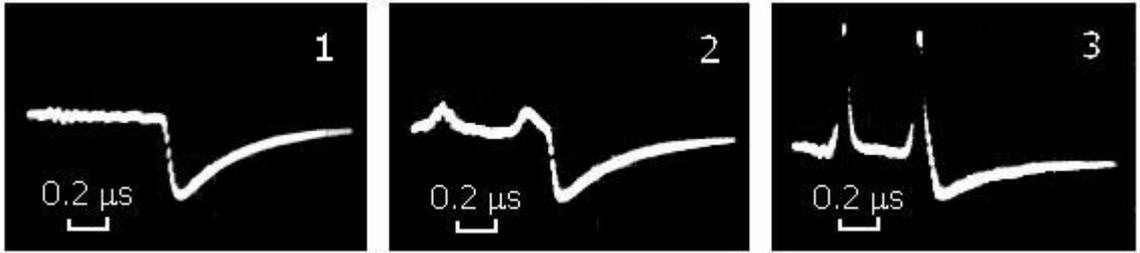

Fig.11. Signals from anode wall probe at discharge interruption in magnetron.
$r_a = 3.2 cm$; $r_c = 1.0 cm$; $L = 7 cm$; $B = 1.9 kG$; $V = 1.5 kV$; $p = 5 \times 10^{-5} Torr$.

From the oscillograms one can determine the contribution of electron sheath and vortex structure to the electric field on the anode surface. The total value of electric field on the anode surface should include as well a vacuum electric field that is easy to calculate. One can also determine the electric field on the drift orbit of vortex structure $E_0 = 2\pi B r_0 / cT$, if the orbit radius and the period of rotation of vortex structure about the axis of discharge device are known. And finally, the photos of electron sheath along the axis of discharge device allow to determine its thickness.

## VII. Conclusion

Nonneutral electron plasma is very sensitive to the external actions. Therefore, one of the most widely used and universal methods of local diagnostics of plasma – electric and magnetic probes appear to be inapplicable for such plasma. Due to this fact, there appeared the necessity for development of nonperturbing experimental methods of investigation for studying the structures and local inhomogeneities of nonneutral electron plasma. Such methods were developed depending on the way of obtaining the nonneutral electron plasma and on its configuration.

A gas-dicharge nonneutral electron plasma exists for an unlimited long time and is self-sustaining medium with internal source of electrons at the expense of ionization. It can be easily obtained in geometries of magnetron or inverted magnetron type, i.e. in the space between two coaxial cylindrical electrodes. The experimental method of investigation of such plasma should not only not perturb it, but not interrupt the discharge as well, in order not to perturb the natural course of process development in plasma. Just for such electron plasma the method of two wall probes described in detail in the present paper has been developed. The method of two wall probes consists in simultaneous measurements of signals from wall probes of the anode and of the cathode during the motion of vortex structure about the axis of discharge device. It allows to follow continuously the trajectory and the charge of one or several vortex structures for a long period of time. In combination with the measurement of electron ejection from the vortex structures, this method gives the possibility to determine the parameters of vortex structures, to investigate their formation, interaction and dynamics. Though this method was developed for geometries of magnetron and inverted magnetron, its modification is applicable as well in the case of Penning cell [10].

In contrast to gas-discharge nonneutral electron plasma, the pure electron plasma is formed by injection of electrons to the trap with crossed electric and magnetic fields. For the most investigations, the Penning-Malmberg trap is used. By its construction this trap is very close to the Penning cell. The both devices are a hollow cylindrical anode located in longitudinal magnetic field and limited by cathodes at the ends: flat as in the Penning cell or in the form of short cylinders as in the Penning-Malmberg trap. However, this difference in the form of cathodes has a very significant importance. In the Penning cell the flat cathodes serve as a source of primary electrons at the expense of ion-electron emission. Therefore, the Penning cell, like the magnetron and the inverted magnetron is used for ignition of discharge in the crossed electric and magnetic fields. In the Penning-Malmberg trap the primary electrons "perish" on the cathodes under the action of magnetic field and ignition of discharge in such geometry is very difficult or even impossible. But on the other hand, the Penning-Malmberg trap is ideally adapted to the external injection of electrons and to their "extraction" from the trap after a certain period of time. Pure electron plasma with the given parameters is injected periodically to the Penning-Malmberg trap, where it "decays" gradually, as the reproduction of electrons does not take place in it. Therefore, the vortex structures in such plasma are formed only at definite initial conditions or are formed artificially. For investigation of vortex structures in such plasma the method of phosphor screen diagnostic was developed [29-31]. This method consists in instantaneous ejection of all electrons from the trap to the luminescent screen along the magnetic field (at first, instead of luminescent screen a radially movable electron collector was used located behind a slit in the endplate). This method gives the spatial, in plane ($r, \theta$), pattern of arrangement and shapes of vortex structures at the given moment of time. However, it is connected with the destruction of plasma imposing the strict conditions on repeatability of the process, as each time we investigate the other structures. For obtaining an exact pattern of the development of the process in time, it is necessary for each cycle to provide not only the identical initial parameters of plasma, but also the appearance of vortex structures in one and the same place, at one and the same time and at one and the same mode of diocotron instability. For this purpose, it is necessary to specify the controllable perturbation of electron plasma.

Thus, the method of phosphor screen diagnostic allows to obtain a full spatial but statistically averaged pattern of the behavior of vortex structures, while the method of two wall probes, on the contrary, shows spatially not full but continuous in time evolution of vortex structures in electron nonneutral plasma. In the Penning-Malmberg trap it is impossible to use the method of two wall probes due to the absence of the central cathode. However, one can use the anode wall probe to obtain the additional information on vortex structures dynamics before destruction of plasma. On the other hand, in gas-discharge nonneutral electron plasma one can use the method of phosphor screen diagnostic to obtain an exact spatial pattern of electron ejection along the magnetic field, e.g. at the moments of passing the inhomogeneities by the wall probes. In the case of discharge interruption we can obtain a full spatial pattern of distribution of electron density in the discharge at the moment of its interruption.

Unfortunately, none of these methods can be used in closed configurations with pure electron plasma, as the closed configurations do not have the open ends for phosphor screen diagnostic and do not have the internal electrode for the method of two wall probes. Therefore, in closed configurations usually the experimental methods are applied in which the emissive Langmuir probes are used [20-23]. In addition, as a nonperturbing diagnostic, sometimes an array of wall probes is used [23], that allows to estimate the location of the charged clumps.